\documentclass[%
 reprint,
 superscriptaddress,
 amsmath,amssymb,
 aps,
 prmaterials,
floatfix,
showkeys
]{revtex4-2}

\usepackage{graphicx}
\usepackage{dcolumn}
\usepackage{bm}
\usepackage[english]{babel}
\usepackage[colorlinks,citecolor=blue]{hyperref}
\usepackage[final]{microtype}
\usepackage{booktabs}
\usepackage{xspace}
\usepackage{tikz}
\usepackage{nicematrix}
\usepackage{miller}
\usepackage{siunitx}
\usepackage{placeins}
\usepackage{lipsum}
\usetikzlibrary{decorations.pathreplacing}

\newcommand{\fig}[1]{Fig.~\ref{#1}\xspace}
\newcommand{\tab}[1]{Tab.~\ref{#1}\xspace}
\newcommand{\app}[1]{Appendix~\ref{#1}\xspace}
\newcommand{\equ}[1]{Eq.~\eqref{#1}\xspace}

\newcommand{\etal}{et al.\xspace}
\renewcommand{\v}{\vec}
\renewcommand{\t}{\mathbf}
\newcommand{\natoms}{\ensuremath{N}\xspace}

\newcommand{\greent}{\ensuremath{\t{   G}}\xspace}        
\newcommand{\stifft}{\ensuremath{\t{\Phi}}\xspace}        
\newcommand{\surfgreent}{\ensuremath{{\greent}_s}\xspace} 
\newcommand{\surfstifft}{\ensuremath{{\stifft}_s}\xspace} 

\newcommand{\surfforcev}{\ensuremath{\v{f}_s}\xspace}     
\newcommand{\surfdisplv}{\ensuremath{\v{u}_s}\xspace}     
\newcommand{\surfstiffcomp}[1]{\ensuremath{\Phi_{s,#1}}\xspace}
\newcommand{\boundaryforce}{\ensuremath{\surfforcev}\xspace}
\newcommand{\boundarydispl}{\ensuremath{\surfdisplv}\xspace}
\newcommand{\boundarygreent}{\ensuremath{\surfgreent}\xspace} 
\newcommand{\boundarystifft}{\ensuremath{\surfstifft}\xspace} 
\newcommand{\contsurfgreent}{\surfgreent} 
\newcommand{\contsurfstifft}{\surfstifft} 
\newcommand{\traction}{\ensuremath{Q}}

\usepackage{xparse}
\newcommand{\realspacecoordinatesymbol}{x}
\newcommand{\recispacecoordinatesymbol}{q}
\newcommand{\realspacecoordvec}{\vec{\realspacecoordinatesymbol}}

\NewDocumentCommand\xc{mo}{%
    \ensuremath{%
        \IfNoValueTF{#2}%
        {\realspacecoordinatesymbol_{#1}}%
        {\realspacecoordinatesymbol_{#1}^{(#2)}} 
    }%
}
\NewDocumentCommand\xcn{mo}{%
    \ensuremath{%
        \IfNoValueTF{#2}%
        {\tilde{\realspacecoordinatesymbol}_{#1}}%
        {\tilde{\realspacecoordinatesymbol}_{#1}^{(#2)}} 
    }%
}
\NewDocumentCommand\xco{mo}{%
    \ensuremath{%
        \IfNoValueTF{#2}%
        {\tilde{s}_{#1}}%
        {\tilde{s}_{#1}^{(#2)}} 
    }%
}
\newcommand{\qc}[1]{\ensuremath{{\recispacecoordinatesymbol_{#1}}}\xspace} 
\newcommand{\xabs}[1]{\ensuremath{\realspacecoordinatesymbol^{(#1)}}} 
\newcommand{\functionxxx}{\ensuremath{\xc{1},\xc{2},\xc{3}{}}}
\newcommand{\functionqqx}{\ensuremath{\qc{1},\qc{2},\xc{3}{}}}
\newcommand{\functionxx}{\ensuremath{\xc{1},\xc{2}}}
\newcommand{\functionqq}{\ensuremath{\qc{1},\qc{2}}}
\newcommand{\pairpot}[1]{\ensuremath{{\phi^{(#1)}}}}
\newcommand{\embedfun}[1]{\ensuremath{{U^{(#1)}}}}
\newcommand{\totedens}[1]{\ensuremath{{\rho^{(#1)}}}}
\newcommand{\edensfun}[1]{\ensuremath{{g^{(#1)}}}}
\newcommand{\hessianterm}[1]{\ensuremath{\prescript{#1  }{}T_{ij}^{(\nu{}\mu{})}}}

\newcommand{\xset}{\ensuremath{
    \left\lbrace\v{\realspacecoordinatesymbol}\right\rbrace}\xspace}
\newcommand{\funcofxset}{\ensuremath{
    \left(\left\lbrace\v{\realspacecoordinatesymbol}\right\rbrace\right)}\xspace}

\usepackage{zebra-goodies}
\usepackage{mathtools}

\begin{document}
\title{Surface lattice Green's functions for high-entropy alloys}

\author{Wolfram G. N\"ohring}
\affiliation{Department of Microsystems Engineering, University of Freiburg, Georges-K\"ohler-Allee 103, 79110 Freiburg, Germany}

\author{Jan Grie\ss er}
\affiliation{Department of Microsystems Engineering, University of Freiburg, Georges-K\"ohler-Allee 103, 79110 Freiburg, Germany}

\author{Patrick Dondl}
\affiliation{Department of Applied Mathematics, University of Freiburg, Hermann-Herder-Str. 10, 79104, Freiburg, Germany}
\affiliation{Cluster of Excellence livMatS, Freiburg Center for Interactive Materials and Bioinspired Technologies, University of Freiburg, Georges-K\"ohler-Allee 105, 79110 Freiburg, Germany}

\author{Lars Pastewka}
\affiliation{Department of Microsystems Engineering, University of Freiburg, Georges-K\"ohler-Allee 103, 79110 Freiburg, Germany}
\affiliation{Cluster of Excellence livMatS, Freiburg Center for Interactive Materials and Bioinspired Technologies, University of Freiburg, Georges-K\"ohler-Allee 105, 79110 Freiburg, Germany}

\date{\today}

\begin{abstract}

We study the surface elastic response of pure Ni, the random alloy FeNiCr and an average FeNiCr alloy in terms of the surface lattice Green's function. We propose a scheme for computing per-site Green's function and study their per-site variations. The average FeNiCr alloys accurate reproduces the mean Green's function of the full random alloy. Variation around this mean is largest near the edge of the surface Brillouin-zone and decays as $q^{-2}$ with wavevector $q$ towards the $\Gamma$-point. We also present expressions for the continuum surface Green's function of anisotropic solids of finite and infinite thickness and show that the atomistic Green's function approaches continuum near the $\Gamma$-point. Our results are a first step towards efficient contact calculations and Peierls-Nabarro type models for dislocations in high-entropy alloys.

\end{abstract}

\maketitle

\section{Introduction}

Atomistic simulations are routinely used to study atomic-scale details 
of elastic or plastic deformation of materials~\cite{tadmor_modeling_2011}. Frequently the number 
of atoms which are needed to resolve the most important details is 
small  in comparison to the number of atoms that must be included 
in the simulation in order to reduce spurious boundary effects. 
One example is the simulation of a dislocation~\cite{bulatov_computer_2006}. 
Here, one may be interested in atoms close to the dislocation
core. However, a large number of atoms around the core must be included
in the simulation to minimize image forces coming from the boundary. 
Another example is the simulation of a half-space subjected to 
surface traction~\cite{johnson_contact_1985}. One may want to focus on atoms near the surface,
but by vitue of St.\ Venant's principle will have to include many sub-surface atoms to simulate a 
substrate of sufficient thickness. 

Fortunately, there are methods for reducing the number of atoms while minimizing 
boundary effects. The general approach is to work in linear response where the relationship between displacements and forces at the boundary can be expressed using a Green's function that captures the sub-surface deformation. This is most straightforward in small-strain elasticity that is linear by construction. Such \emph{continuum} approaches  rely either on a discretization of the continuum domain (e.g.\ by finite 
elements -- see for example Refs.~\cite{kohlhoff_crack_1991, tadmor_quasicontinuum_1996,shenoy_quasicontinuum_1998,xiao_bridging_2004,badia_force-based_2007,chen_concurrent_2019}) or through analytical or semianalytical Green's functions (e.g. Refs.~\cite{amba-rao_fourier_1969,kalker_minimum_1972,stanley_fft-based_1997,hodapp_lattice_2019}). Here, we focus on a class of methods where the atomistic domain 
is coupled to a flexible \emph{atomistic} boundary governed by an elastic lattice Green's function, see e.g.\ Refs.~\cite{sinclair_improved_1971,sinclair_flexible_1978,gallego_harmonic_1993,li_efficient_2009,campana_practical_2006,trinkle_lattice_2008,pastewka_seamless_2012}. The use of lattice Green's function allows to formulate a coupling scheme that is seamless since it can be formulated within a single Hamiltonian~\cite{pastewka_seamless_2012} and hence does not give rise to ghost forces~\cite{miller_unified_2009}.

\begin{figure}[htb!]
    \centering
    \includegraphics{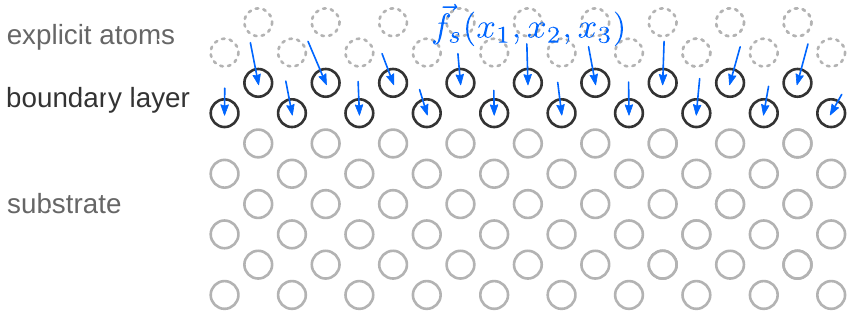}
    \caption{\label{fig:gfmd}Illustration of a case
    where a Green's function method can be used 
    to reduce the number of atoms that need to be 
    included in the simulation; a half-space is 
    partitioned into boundary atoms (dark) and 
    substrate atoms (light); the net force on the
    latter is zero; the displacement of boundary 
    atoms due to a force field 
    $\boundaryforce(\functionxxx)$ (coming from substrate atoms and 
    optional explicit atoms) can be calculated 
    without simulating substrate atoms 
    if the Green's functions \boundarygreent 
    (which depends on substrate conditions, however) 
    is known; conversely, the force on boundary atoms
    can be calculated from their displacements if 
    the effective stiffness \boundarystifft is known.
    }
\end{figure}

In the Green's function method described in Refs.~\cite{campana_practical_2006,pastewka_seamless_2012}, the response of the removed substrate atoms is approximated by modifying the forces on atoms in the boundary region, see \fig{fig:gfmd}. The forces on atoms in this region are given by an effective stiffness tensor, the elastic Green's function of the boundary layer. Refs.~\cite{campana_practical_2006,pastewka_seamless_2012} and the present work only regard the static limit, but the general approach outlined below is also amenable to a dynamic treatment~\cite{kajita_simulation_2012,pastewka_seamless_2012,kajita_greens_2016,joseph_m_monti_greens_nodate}.

Let $\boundaryforce(\functionxxx)$ be the (static) force on the atoms in the boundary
region as a function of coordinates $\xc{1}$ and $\xc{2}$ in the plane,
and coordinate $\xc{3}$ perpendicular to the plane with positive $\xc{3}$ pointing into the substrate. Taking the Fourier
transform with respect to the in-plane coordinates yields
the representation $\boundaryforce(\functionqqx)$, with wavenumbers
$\qc{1}$ and $\qc{2}$. If no external forces act on the 
substrate atoms, and $\boundaryforce(\functionqqx)$ were known,
then in the static limit the displacements $\boundarydispl(\functionqqx)$ in the boundary
could be calculated using the elastic Green's function $\boundarygreent(\functionqqx)$, 
\begin{align}
    \boundarydispl(\functionqqx) = \boundarygreent(\functionqqx)  \cdot \boundaryforce(\functionqqx). 
    \label{eq:greenft}
\end{align}
Inversion yields 
\begin{align}
    \boundaryforce(\functionqqx) = \boundarystifft(\functionqqx)  \cdot \boundarydispl(\functionqqx), 
    \label{equ:stiffnessdef}
\end{align}
where $\boundarystifft={\boundarygreent}^{-1}$ is a matrix of complex stiffness
coefficients, which depends on the substrate configuration. The problem
has thus been shifted from simulating substrate atoms to determining 
\boundarystifft. In the case  of unary systems, \boundarygreent and 
\boundarystifft have been measured in molecular dynamics simulations using a fluctuation-dissipation 
theorem \cite{campana_practical_2006,kong_implementation_2009}, or directly calculated using a 
transfer matrix or renormalization group approach \cite{pastewka_seamless_2012}. 

The outlined method is limited to homogeneous crystals, since it relies on the assumption that  the elastic constants of the medium are invariant under translation. This assumption breaks down in alloys. As a first step towards extending this method for alloys, we examine in this paper the variation of the surface stiffness in a multi-principal 
element random alloy, where every atom has a random environment. The surface stiffness is 
the special case of  $\boundarystifft$ where only surface atoms are retained.

We used the Embedded Atom Method (EAM) \cite{Daw_1984} and calculated $\surfstifft$ by inversion of the Hessian matrix $\mathbf{H}$ of the potential energy, the analytical solution of which we have derived for this class of interatomic potentials. We compared the stiffness of the true random alloy to the stiffness of a mean field model of the alloy, where we used the Average-atom (A-atom) \cite{Varvenne_2016} approximation. Additionally, we present the anisotropic-elastic solution for $\surfstifft$ of a continuous half-space with finite thickness and compare the continuum stiffness to atomistic data. 

Our results show that the atomistic solutions converges to the continuum in the limit of long wavelengths. At short wavelengths, the local environment of atoms controls $\surfstifft$, hence the continuum solution  is a poor estimate. The average alloy model is a fair approximation for the mean stiffness of the true random alloy at all wavelengths, but fluctuations grow substantially as the wavelength decreases.

\section{Methods}

\subsection{Atomistic stiffness}
\begin{figure}[htb!]
    \centering
    \includegraphics{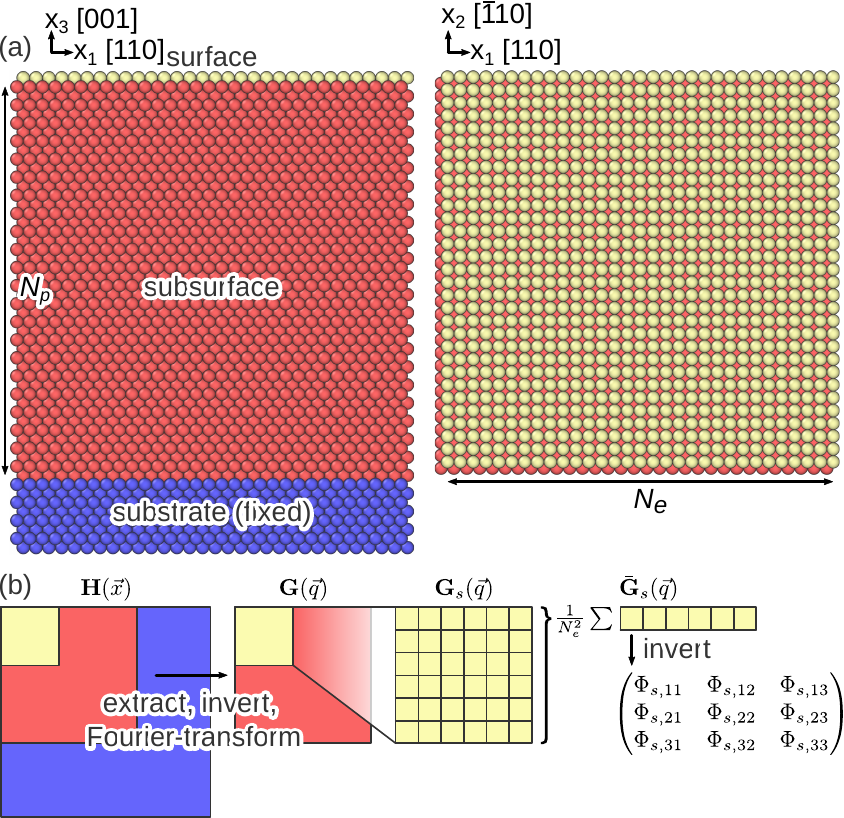}
    \caption{(a) Configuration used for calculating the surface stiffness $\surfstifft$ in side view (left) and top view (right). Atoms in the substrate layers (blue, thickness two times the potential cutoff radius) were fixed in positions corresponding to static equilibrium under fully periodic boundary conditions. Atoms in the subsurface (red) and surface (yellow) layers were free. Periodic boundary conditions were applied in the $\xc{1}$- and $\xc{2}$-directions. The $\hkl(001)$ planes (perpendicular to the $\xc{3}$-direction) are square lattices
    with $N_e$ atoms along the edges; the subsurface region consists of $N_p$ 
    $\hkl(001)$ planes. 
    (b) Illustration of the calculation of $\surfstifft$;
    the upper left block of the Hessian matrix $\t{H}$ (left) involves only 
    surface and subsurface atoms; this block is inverted and Fourier-transformed to obtain Green's functions $\greent$ (middle); the upper left  $3N_e\times{}3N_e$ sub-block corresponding to interactions between surface atoms contains the surface Green's functions $\surfgreent$; each of the $N_e^2$ $3\times3$ blocks contains the values of Green's functions for one pair of atoms $\nu\mu$, i.e.\ one wavevector $(\qc{1},\qc{2})$ in the plane; 
    one row of blocks corresponds to the  Green's functions with one 
    site $\nu$ fixed; 
    the surface stiffness matrices $\surfstifft$ are the harmonic averages obtained by first averaging block rows of $\surfgreent$ and then calculating the matrix inverse of each block
    \label{fig:atomic_config}}
\end{figure}

We calculated the stiffnesses of the \hkl(001) surface of face-centered cubic (FCC) crystals using square slab configurations, see \fig{fig:atomic_config}. The simulation cell was a rectangular prism with a square base. Consider a Cartesian coordinate system  with directions $\xc{1}$ and $\xc{2}$ in the plane of the base. In order to simulate a half-space, we applied periodic boundary conditions along the $\xc{1}$- and $\xc{2}$-directions, and open boundary conditions along the $\xc{3}$-direction. The lattice directions \hkl[110], \hkl[1-10],  and \hkl[001] of the crystal were parallel to the $\xc{1}$-, $\xc{2}$-, and $\xc{3}$-directions of the cell.

The set of \hkl(001) planes can be partitioned into surface, subsurface, and substrate planes. There is only one surface plane, but $N_p$ subsurface and $N_u$ substrate planes. We call the corresponding sets of atoms $\Omega_s$, $\Omega_p$ and $\Omega_u$. Atoms in \hkl(001) planes form a 2D square lattice with  lattice parameter $a=a_0/\sqrt{2}$, 
where $a_0$ is the FCC lattice parameter.  
Let there be $N_e$ atoms along the edge in the $\xc{1}$- or $\xc{2}$-direction, then there are $N_e^2$ atoms in each plane, and the total number of atoms is $N=(1+N_p+N_s)N_e^2$. 
In order to eliminate surface effects at the bottom boundary, we fixed the substrate atoms in positions corresponding to static equilibrium under fully periodic boundary conditions and made the fixed substrate layer thicker than the maximum interaction distance of the potential. We used Embedded Atom Method (EAM) \cite{Daw_1984} potentials, hence the required thickness was two times the cutoff radius $r_\textrm{cut}$ of the potential. The process of constructing the configurations depends on the material and is explained in detail below.

To determine the surface stiffnesses $\surfstifft$, we first calculated the Green's functions $\greent$ by inverting the Hessian matrix $\t{H}$ of the potential energy $\mathcal{V}^\mathrm{int}$, see \fig{fig:atomic_config}(b). The components of $\t{H}$ are
\begin{align}
H_{(3(\nu{}-1)+i)(3(\mu{}-1)+j)}
=\frac{\partial^2 \mathcal{V}^\mathrm{int} \funcofxset}
      {\partial \xc{i}[\nu] \partial \xc{j}[\mu] }\bigg\vert_{\xset_0},
\end{align}
where indices $\mu, \nu \in [1,\natoms]$ refer to atoms, and indices 
$i,j\in[1,2,3]$ refer to the three components of a vector in $\xc{1}$-,
$\xc{2}$- and $\xc{3}$-direction. In this equation $\xset$ is the set of 
position vectors of the atoms $\xset\equiv\left\lbrace{\realspacecoordvec}^{(1)},\dots,{\realspacecoordvec}^{(N)}\right\rbrace$,
and $\xc{i}[\nu]$ is the coordinate of atom 
$\nu$ in $i$-direction. $\xset_0$ is the set of equilibrium positions where the force on the atoms vanishes.  We used the analytical 
solution of $\t{H}$ for EAM potentials, see the derivation in \app{app:hessian}. This solution is implemented in the Python package \texttt{matscipy} \cite{matscipy}.

$\t{H}$ is a real symmetric $3\natoms\times{}3\natoms$ matrix.
It is sparse, because the range of interaction between
atoms is limited. Consider an arbitrary displacement of the atoms, written as a $3N$-dimensional vector $\vec{u}$. Within the harmonic approximation, the components of the resulting 
force vector $\v{f}$ are 
\begin{align}
    f_{3(\nu-1)+i}= H_{((3\nu-1)+i)(3(\mu-1)+j)}u_{3(\mu-1)+j}, 
\end{align}
where repeated indices imply summation over the corresponding range. Recall that the atoms in 
$\Omega_{u}$ are fixed in their equilibrium 
positions, hence $u$ vanishes for those atoms. To impose this constraint,
we computed $\t{H}$ of the whole configuration,
but then eliminated elements in $\t{H}$ corresponding
to pairs of atoms where one or both of them are in 
$\Omega_u$. The remaining $3(1+N_p)N_e\times 3(1+N_p)N_e$ elements correspond to pairs of atoms in $\Omega_s\cup\Omega_p$. It is convenient to label the atoms such 
that these elements form the upper left block of $\t{H}$ (see Fig.~\ref{fig:atomic_config}b).

This block was then inverted to obtain Green's functions 
$G_{(3(\nu-1)+i)(3(\mu-1)+j)}$, which solve
\begin{align}
    u_{3(\nu-1)+i} = G_{(3(\nu-1)+i)(3(\mu-1)+j)}f_{3(\mu-1)+j}, 
    \label{equ:greensubscript}
\end{align}
where $\nu,\mu\in{}\Omega_s\cup\Omega_p$. To bring out the block structure
of $G_{3(\nu-1)+i)(3(\mu-1)+j)}$, we can rewrite this equation as 
\begin{align}
     \v{u}^{(\nu)}
    = \sum_{\mu\in{}\Omega_s\cup\Omega_p} 
      \t{G}^{(\nu,\mu)} \cdot
    \vec{f}^{(\mu)}, \label{equ:greenrecast}
\end{align}
where $\v{u}^{(\nu)}$ is the displacement vector of atom $\nu$, with components $u_{3(\nu-1)+i}$ ($i=1,2,3$) in \equ{equ:greensubscript}; 
$\t{G}^{(\nu,\mu)}$ is the $3\times{}3$ block $G_{(3(\nu-1)+i)(3(\mu-1)+j)}$; 
and $\v{f}^{(\mu)}$  corresponds to the components $f_{3(\mu-1)+i}$ ($i=1,2,3$).

We inverted $\t{H}$ via Cholesky factorization, using \textsc{petsc}  \citep{petsc-user-ref,petsc-efficient}
and \textsc{mumps} \cite{MUMPS:1,MUMPS:2}. The latter allows parallel 
calculation of the selected entries \cite{amestoy_parallel_2015} in
the upper left block of the matrix.
For pure crystals, the surface Green's function can be computed efficiently using renormalization group approaches for system sizes beyond billions of atoms~\cite{pastewka_seamless_2012}.

Using the same argument to eliminate the substrate atoms from $\t{H}$, we now eliminate the subsurface atoms from $\t{G}$: the forces on the subsurface atoms must vanish since we are working in the static limit -- which means that all subsurface atoms remain in their equilibrium positions. The remaining quantity is a $3N_s\times{}3N_s$ matrix $\surfgreent$ describing the degrees of freedom corresponding to the
surface atoms. \equ{equ:greenrecast} (with $\nu,\mu\in\Omega_s$) can be 
interpreted as a signal measured at $3N_s\times{}3N_s$ points $\{(\xc{1}[\nu],\xc{2}[\nu])\mid{}\nu\in\Omega_s\}$ in the plane, 
and we write 
\begin{align}
    \v{u}_s(\xc{1}[\nu],\xc{2}[\nu])
    = \sum_{\mu\in{}\Omega_s} 
      \surfgreent(\xc{1}[\nu],\xc{2}[\nu],\xc{1}[\mu],\xc{2}[\mu]) \cdot
    \v{f}_s(\xc{1}[\mu],\xc{2}[\mu{}]). \label{equ:greendiscrete}
\end{align}

In a pure crystal with translational symmetry in the plane, Green's functions would only 
depend on the relative distance between points, i.e.\ 
$\surfgreent(\xc{1}[\nu],\xc{2}[\nu],\xc{1}[\mu],\xc{2}[\mu]) \rightarrow   
 \surfgreent(\xc{1}[\nu]-\xc{1}[\mu],\xc{2}[\nu]-\xc{2}[\mu])$, and 
\equ{equ:greendiscrete} would be a convolution. According
to the convolution theorem, taking the discrete Fourier 
transform of \equ{equ:greendiscrete} would then produce
\equ{equ:stiffnessdef}.

However, in random alloys translational symmetry is
broken. Thus, we studied the variation of 
stiffness across sites. We denote the Green's functions
\emph{of site} $\nu$ as 
\begin{equation}
    \surfgreent^{(\nu)}(\xc{1}[\nu]-\xc{1}[\mu],\xc{2}[\nu]-\xc{2}[\mu])
    \equiv{}
    \surfgreent(\xc{1}[\nu],\xc{2}[\nu],\xc{1}[\mu],\xc{2}[\mu])
    \label{equ:greenrel}
\end{equation}
in what follows. Note that in \equ{equ:greenrel} the argument of $\surfgreent^{(\nu)}$ is measured with respect to the position of the site $\v{x}^{(\nu)}$ such that we can carry out a Fourier-transform for each site. This representation is useful for comparison with the unary system and the continuum solution, where the per-site variation disappears from \equ{equ:greenrel}. Neglecting non-affine microdistortions~\cite{song_local_2017,owen_lattice_2018,owen_quantifying_2020}, the atoms are arranged approximately in a simple cubic lattice within the periodic domain. Hence in equilibrium
\begin{align}
\begin{split}
    (\xc{1}[\nu]-\xc{1}[\mu],\xc{2}[\nu]-\xc{2}[\mu]) &\approx (m a, n a), \\ &\text{with}\;m,n\in\left\lbrace 0,\dots,N_e-1\right\rbrace.
\end{split}
\end{align}
The discrete Fourier transform of $\surfgreent^{(\nu)}\left(m a,n a\right)$ is 
\begin{widetext}
\begin{align}
\begin{aligned}
    \surfgreent^{(\nu)}\left(\frac{2\pi}{aN_e}k, \frac{2\pi}{aN_e}l \right)= 
    &\sum_m^{N_e-1} 
    \sum_n^{N_e-1}
    \surfgreent^{(\nu)}\left(m a,n a\right) 
    \exp\left\lbrace
        -2\pi{}i\left(\frac{m k}{N_e}+\frac{m l}{N_e}\right)
    \right\rbrace,\;\text{with}
    \;k,l\in\left\lbrace 0,\dots,N_e-1\right\rbrace
\end{aligned}
\end{align}.
\end{widetext}
Thus, we map the 
solution to $N_e\times{}N_e$ wavevectors in the 
first Brillouin zone with components
\begin{align}
    \qc{1},\qc{2}\in
    \begin{dcases}
        \left\lbrace 
            -\frac{N_e}{2},\dots,\frac{N_e}{2}-1
        \right\rbrace \cdot  \frac{2\pi }{aN_e}
      &\, \text{$N_e$ even}  \\ 
      \left\lbrace 
            -\frac{N_e-1}{2},\dots,\frac{N_e-1}{2}
        \right\rbrace \cdot  \frac{2\pi }{aN_e}
      &\, \text{$N_e$ odd}
    \end{dcases}.
\end{align}
However, due to symmetry only the quadrant $0\leq{}\qc{1},\qc{2}\leq{\pi/a}$ is unique. 

In an unary system, $\surfgreent^{(\nu)}(\functionqq)$ is the same for all sites $\nu$. In a random alloy, on the other hand, there are site-by-site variations, 
and the components $G_{s,ij}^{(\nu)}(\functionqq)$ of $\surfgreent^{(\nu)}(\functionqq)$ become complex random variables.

Rather than discussing the surface Greens function itself $\surfgreent(\v{q})$, it is convenient to discuss its inverse, the surface stiffness $\surfstifft(\v{q})$. This is because in the long-wavelength (continuum) limit $\surfstifft(\v{q}) \propto q$~\cite{amba-rao_fourier_1969,campana_practical_2006,pastewka_seamless_2012} while $\surfgreent(\v{q})$ diverges. The question we will discuss in the following is how to characterize the mean response of the solid and the magnitude of per-site fluctuations.

We first computed a mean stiffness  $\surfstifft(\functionqq)$. Consider the related problem of 
stochastic homogenization of an elastic continuum with 
fluctuating elastic constants. Here,
the effective stiffness of the homogenized medium 
can be computed using the solution of a corrector equation \cite{MR542557, MR712714}. The solutions of the equation with random coefficients then converge to the solution of the homogenized equation in mean.
Let 
\begin{align}
\surfstifft^{(\nu)}(\functionqq)\equiv\left(\surfgreent^{(\nu)}(\functionqq)\right)^{-1}
\end{align} be the stiffness of site $\nu$. 
Following the continuum approach, we computed the average surface stiffness as 
\begin{align}
    \begin{aligned}
        \surfstifft(\functionqq)  = \left\langle\left(\surfstifft^{(\nu)}(\functionqq)\right)^{-1}\right\rangle^{-1} = {\bar{\t{G}}_s}^{-1}\left(\functionqq\right), \label{equ:harmonicavg}
    \end{aligned}
\end{align}
where $\left\langle\dots\right\rangle$ indicates the arithmetic 
mean over all sites, i.e.\ the components of ${\bar{\t{G}}_s}\left(\functionqq\right)$
are 
\begin{align}
    \bar{G}_{s,ij}(\functionqq) = \frac{1}{N_e^2} \sum_{\nu=1}^{N_e^2} G_{s,ij}^{(\nu)}(\functionqq).  \label{equ:arithmeticavggreen}
\end{align}

We calculated $\surfstifft(\functionqq)$  of  pure Ni, and $\surfstifft(\functionqq)$   of a random solid solution of Fe, Ni and Cr with equal concentration of all elements. In both cases, we used the EAM potential by Bonny \etal \cite{bonny_interatomic_2011}, which has a smooth cutoff with continuous first and second derivatives. We prepared the random Fe-Ni-Cr alloy by 
randomly assigning the constituent elements to the lattice sites. 
Additionally, we performed calculations with a mean-field model. Here, we replaced the different real elements by a single ``average'' element, the A-atom \cite{Smith_1991,Varvenne_2016}.  It behaves like a pure metal with similar average properties as the true random solution. A module for generating A-atom potentials has been implemented in the Python package \texttt{matscipy} \cite{matscipy}. In pure Ni and the A-atom crystal, $\surfstifft^{(\nu)}(\functionqq)$ is the same for all surface sites $\nu$, 
hence $\surfstifft(\functionqq)=\surfstifft^{(\nu)}(\functionqq)$. The 
harmonic mean of the random alloy data according to \equ{equ:harmonicavg} can
be compared to the A-atom solution and the continuum solution. The latter
requires only three cubic elastic constants as input. 

In the case of the Ni and A-atom configurations, our starting point was a perfect crystal lattice with the appropriate \SI{0}{\kelvin} lattice parameter. We then minimized the potential energy of the surface and subsurface atoms using \textsc{fire} \cite{bitzek_structural_2006,guenoele_assessment_2020}.  The substrate atoms were fixed during minimization and the iteration  was stopped as soon as the Euclidean norm of the global force 
vector fell below \SI{1e-6}{\electronvolt\per\angstrom}. In order to create the random FeNiCr alloy, we started from a thicker, fully periodic configuration. We added an additional slab of atoms of thickness greater than  $2r_\textrm{cut}$ below the substrate and then minimized the potential energy  under fully periodic boundary conditions. This minimization allowed the substrate atoms to move to their non-affine equilibrium positions in the bulk. Afterwards, we opened the  boundary along the $\xc{3}$-direction and removed the extra atoms. Finally, we minimized the potential energy of the open system with fixed substrate atoms. 

The required lattice parameters and elastic constants of the three materials are listed in  \tab{tab:material_properties} in \app{app:propertycalc}. See this 
appendix also for details on how the values were computed. 

\subsection{Continuum stiffness}

\begin{figure}[htb!]
    \centering
    \includegraphics{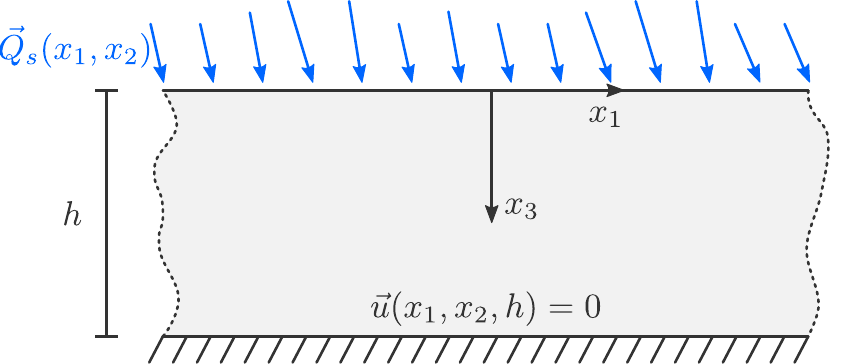}    
    \caption{Model used for deriving a continuum solution of 
    the surface Green function and the associated stiffness
    tensor; the surface at $\xc{3}=0$ is subjected
    to tractions $\v{\traction}_s(\xc{1},\xc{2})$; we consider  
    finite thickness
    $h$ with fixed boundary conditions at $\xc{3}=h$, and 
     the limit $h\rightarrow\infty$; 
    the solid is infinite along $\xc{1}$ and $\xc{2}$;
    the material has anisotropic elastic constants $C_{ijkl}$, 
    which are constant throughout the body}
    \label{fig:continuum_model}
\end{figure}

We consider a semi-infinite solid, see \fig{fig:continuum_model}. The body has a surface
perpendicular to the $\xc{3}$-direction and extends to infinity in the $\xc{1}$- 
and $\xc{2}$- directions. Positive $\xc{3}$ are located within the solid. In the $\xc{3}$-direction, the body may have a finite thickness
$h$, or infinite thickness $h\rightarrow\infty$. We discuss both cases. 
Tractions $\vec{\traction}(\functionxx)$ are applied at the surface and there are no body 
forces. In the case where the body has finite thickness $h$, we assume a fixed boundary, i.e.\ $\vec{u}(\xc{1},\xc{2},h)=0$. The material is homogeneous and linear-elastic, with anisotropic elastic constants $C_{ijkl}$ ($i,j,k,l \in [1,2,3]$) subject to the usual symmetry requirements~\cite{barber_elasticity_1992}. For the FCC solid considered here, there are three indendent elastic constant that are typically denoted by $C_{11}$, $C_{12}$ and $C_{44}$.

We are interested in the surface displacements $\vec{u}_s(\functionxx)$ in elastostatic equilibrium, where the divergence of the stress tensor $\sigma$ vanishes, 
\begin{align}
    \partial_i\sigma_{ij} = 0.
    \label{eq:elastoequ}
\end{align}
$\partial_i$ indicates the partial derivative in direction $i$ and Einstein summation convention applies. Eq.~\eqref{eq:elastoequ} corresponds to requiring zero forces for the subsurface atoms in our atomistic calculations.

$\vec{u}_s(\functionxx)$ can be calculated by a convolution of the 
tractions  $\vec{\traction}(\functionxx)$ with surface Green's function 
$\contsurfgreent(\functionxx)$, 
\begin{align}
    \v{u}_s(\functionxx)=\int d\xc{1}'d\xc{2}' \contsurfgreent(\xc{1}-\xc{1}', \xc{2}-\xc{2}')\cdot\v{\traction}(\xc{1}',\xc{2}'). \label{eq:greenconvo}
\end{align}
Fourier transformation yields \equ{eq:greenft} (with $\xc{3}=0$ dropped). 
In Appendix~\ref{app:conti}, we derive $\contsurfgreent(\functionqq)$ for
finite $h$ and $h\rightarrow\infty$. The solution can be 
represented as a matrix product 
\begin{align}
    \contsurfgreent(\functionqq)=\t{U}(\functionqq,0)\cdot\t{F}^{-1}(\functionqq).
\end{align}
The matrices $\t{U}(\functionqq,0)$ and $\t{F}(\functionqq)$ 
depend on the eigenvalues of the Fourier 
transform of the linear operator 
\begin{align}
    M_{il} = C_{ijkl}\partial_j\partial_k,
    \label{eq:Mopdeffirst}
\end{align}
and the admissible basis functions of the displacement field.
No closed-form solution exists, but it is straightforward to 
calculate said eigenvalues and basis functions numerically. 
The inverse of $\contsurfgreent(\functionqq)$ is the surface stiffness 
$\contsurfstifft(\functionqq)$. We have implemented the numerical 
solution of $\contsurfstifft(\functionqq)$ in the Python package
\texttt{ContactMechanics} \cite{pyco}.

\section{Results}
We first calculated the surface stiffness of Ni to obtain reference data for a 
pure metal. The atomic configuration had $N_e=31$ atoms along the edge, and $N_p=45$ subsurface planes. All $N_e^2=961$ solutions of 
$\surfstifft^{(\nu)}$ for different surface sites $\nu$ are equal due to translational 
symmetry in the plane. Notice that $\surfforcev^{(\nu)}$ in \equ{equ:greendiscrete} of the atomistic solution is a \emph{force}, whereas $\v{\traction}$ in \equ{eq:greenconvo} of the continuum solution is a \emph{traction} (units of \SI{}{\newton\per\meter\squared}).
In order to compare atomistic and continuum Green's functions, we need to divide the former by the mean area per atom $a^2=a_0^2/2$. The resulting Green's function has SI units of \SI{}{\newton\per\meter\cubed}, but it is more convenient to use \SI{}{\giga\pascal\per\angstrom}, since $\surfstiffcomp{11}$, for example, should converge to $C_{44}/h$ in the limit of infinite wavelength.

\begin{figure}[hbt!]
    \centering
    \includegraphics{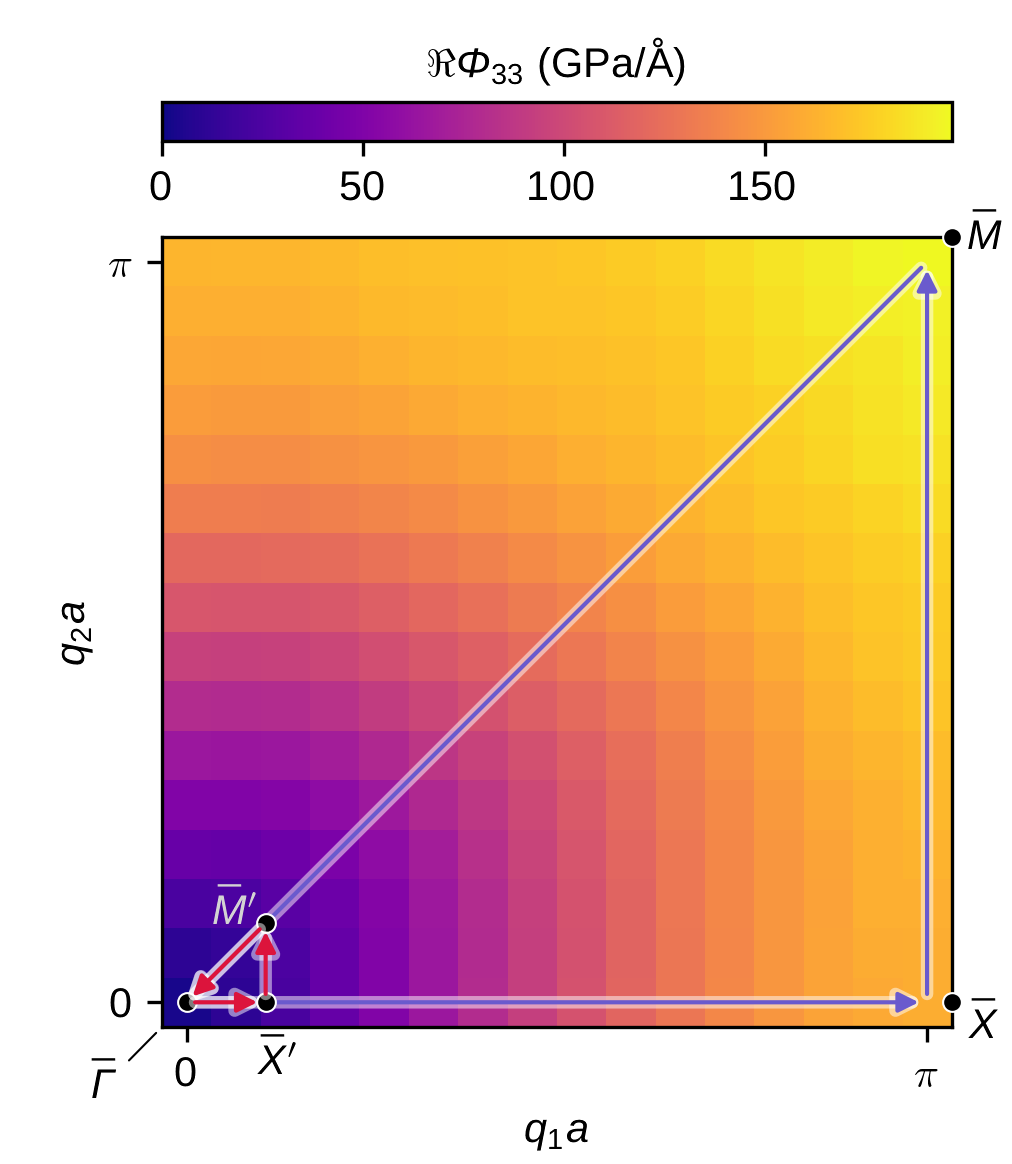}
    \caption{Real part of stiffness component $\surfstiffcomp{33}$ of Ni in a quadrant of the surface Brillouin zone; in other figures we have plotted the values 
    along the paths $\bar{\Gamma}$-$\bar{X}$-$\bar{M}$-$\bar{\Gamma}$ and
    $\bar{\Gamma}$-$\bar{X}'$-$\bar{M}'$-$\bar{\Gamma}$.}
    \label{fig:brillouin} 
\end{figure}

Figure~\ref{fig:brillouin} shows $\Re\surfstiffcomp{33}(\functionqq)$ in 
the quadrant of the surface Brillouin zone where $0\leq{}\functionqq\leq{\pi/a}$. Here and in the following, $\Re$ and $\Im$ 
refer to the real and imaginary parts of an imaginary number, respectively. 
The other  quadrants are symmetric with respect to the $\qc{1}$- and $\qc{2}$-axes. 
The stiffness is minimal in the long-wavelength limit $\qc{1}=\qc{2}=0$ and increases with decreasing wavelength. The origin $\qc{1}=\qc{2}=0$ is called $\bar{\Gamma}$. The center of the edge of the Brillouin zone along $\qc{1}$ is called $\bar{X}$, and the corner is called $\bar{M}$. Below, we show plots of values along the path $\bar{\Gamma}$-$\bar{X}$-$\bar{M}$-$\bar{\Gamma}$. To study convergence in the long-wavelength
limit, we also consider a shorter path $\bar{\Gamma}$-$\bar{X}'$-$\bar{M}'$-$\bar{\Gamma}$,
where $\bar{X}'$ is on the line $\qc{2}=0$, at \SI{10}{\percent} of the distance 
from $\bar{\Gamma}$ to $\bar{X}$. Point $\bar{M}'$ corresponds to a 
wavelength of \SI{11}{\angstrom} along the $\xc{1}$- and $\xc{2}$-directions.

\begin{figure*}[t]
    \centering
    \includegraphics{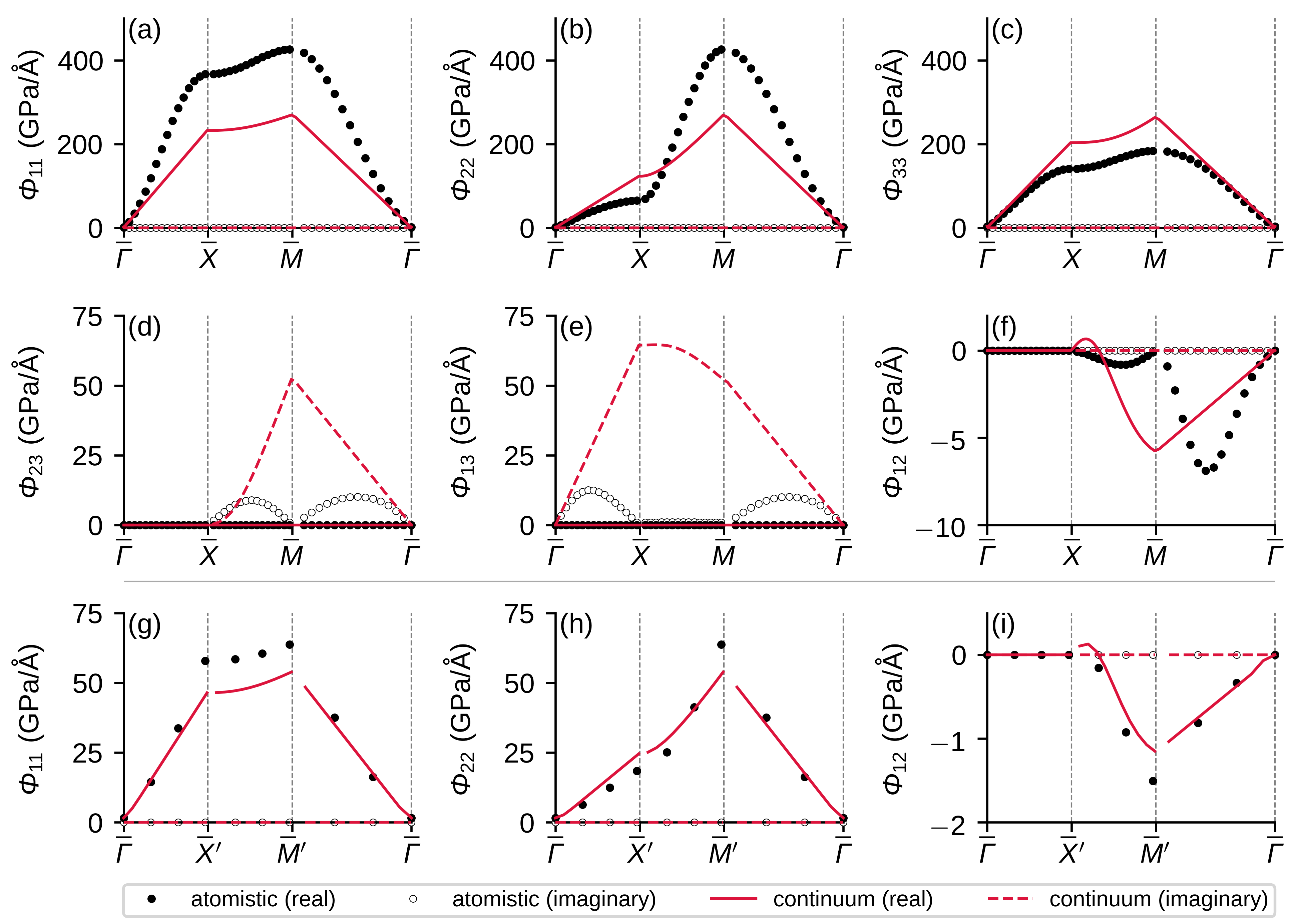}    
    \caption{
    Surface stiffness of Ni; markers: atomistic model; lines: anisotropic-elastic model; the upper two rows show the variation of 
    the six independent components along the path $\bar{\Gamma}$-$\bar{X}$-$\bar{M}$-$\bar{\Gamma}$
    through the first quadrant of the surface Brillouin zone (see \fig{fig:brillouin}); the normal moduli $\surfstiffcomp{11}$, $\surfstiffcomp{22}$, and
    $\surfstiffcomp{33}$, as well as the shear modulus $\surfstiffcomp{12}$ are purely real;
    the two in-plane shear moduli $\surfstiffcomp{23}$ and $\surfstiffcomp{13}$ are purely
    imaginary; atomistic and continuum models disagree near points $\bar{X}$ and
    $\bar{M}$, which are short wavelength limits; the lower row shows the values
    of $\surfstiffcomp{11}$, $\surfstiffcomp{22}$, and $\surfstiffcomp{12}$ along the shorter path $\bar{\Gamma}$-$\bar{X}'$-$\bar{M}'$-$\bar{\Gamma}$ through the long wavelength region; 
    the relative difference between atomistic and continuum models decreases with increasing wavelength
    }
    \label{fig:result_pure}
\end{figure*}

The upper two rows of \fig{fig:result_pure} show the six independent components 
of $\surfstifft$ along the path $\bar{\Gamma}$-$\bar{X}$-$\bar{M}$-$\bar{\Gamma}$. Four moduli are 
purely real, namely the normal moduli  $\surfstiffcomp{11}$, $\surfstiffcomp{22}$, and $\surfstiffcomp{33}$, as well as the in-plane shear modulus $\surfstiffcomp{12}$. The out-of-plane shear moduli $\surfstiffcomp{23}$ and $\surfstiffcomp{13}$ are purely imaginary.
The discrepancy between the continuum data and the atomistic data
increases as one moves away from the long-wavelength limit near $\bar{\Gamma}$
towards $\bar{X}$ or $\bar{M}$. The difference is maximum at corner of the surface Brillouin zone 
$\bar{M}$, which represents the short-wavelength limit in both in-plane 
directions. In the case of the shear moduli, the continuum model fails to  
predict the extrema between $\bar{X}$ and $\bar{M}$ (zone edge), 
and between $\bar{M}$ and $\bar{\Gamma}$ (zone diagonal). The bottom row of 
\fig{fig:result_pure} shows the components  $\surfstiffcomp{11}$, $\surfstiffcomp{22}$, and $\surfstiffcomp{12}$ along the path  $\bar{\Gamma}$-$\bar{X}'$-$\bar{M}'$-$\bar{\Gamma}$.
These plots indicate that the continuum and atomistic solutions
converge near $\bar{\Gamma}$. The solutions for $\surfstiffcomp{12}$ agree qualitatively.  
The atomistic value of $\surfstiffcomp{11}$ and $\surfstiffcomp{22}$ at $\bar{\Gamma}$ is \SI{1.6}{\giga\pascal\per\angstrom}, which is equal to $C_{44}/h$ with $C_{44}=\SI{125}{\giga\pascal}$ and $h=45a_0/2=\SI{79.2}{\angstrom}$.

Having examined the surface stiffness of pure Ni, we now attend to the
alloy case. We created $500$ random alloy samples and one average alloy
sample with $N_p=45$ and $N_e=31$. Below, we  report the \emph{arithmetic} mean of $\Phi_{s,ij}(\functionqq)$ of all random alloy samples. However, sample-by-sample variations of $\Phi_{s,ij}(\functionqq)$ are small, since this is the homogenized stiffness with fluctuations averaged out. Additionally, we  quantified \emph{site-by-site} fluctuations by calculating the 10th and 90th percentiles of the site-specific stiffness tensors $\surfstifft^{(\nu)}(\functionqq)$ at
all sites in all samples, i.e.\ $500\cdot{}N_e^2=480500$ sites in total.

\begin{figure*}[t]
    \centering
    \includegraphics{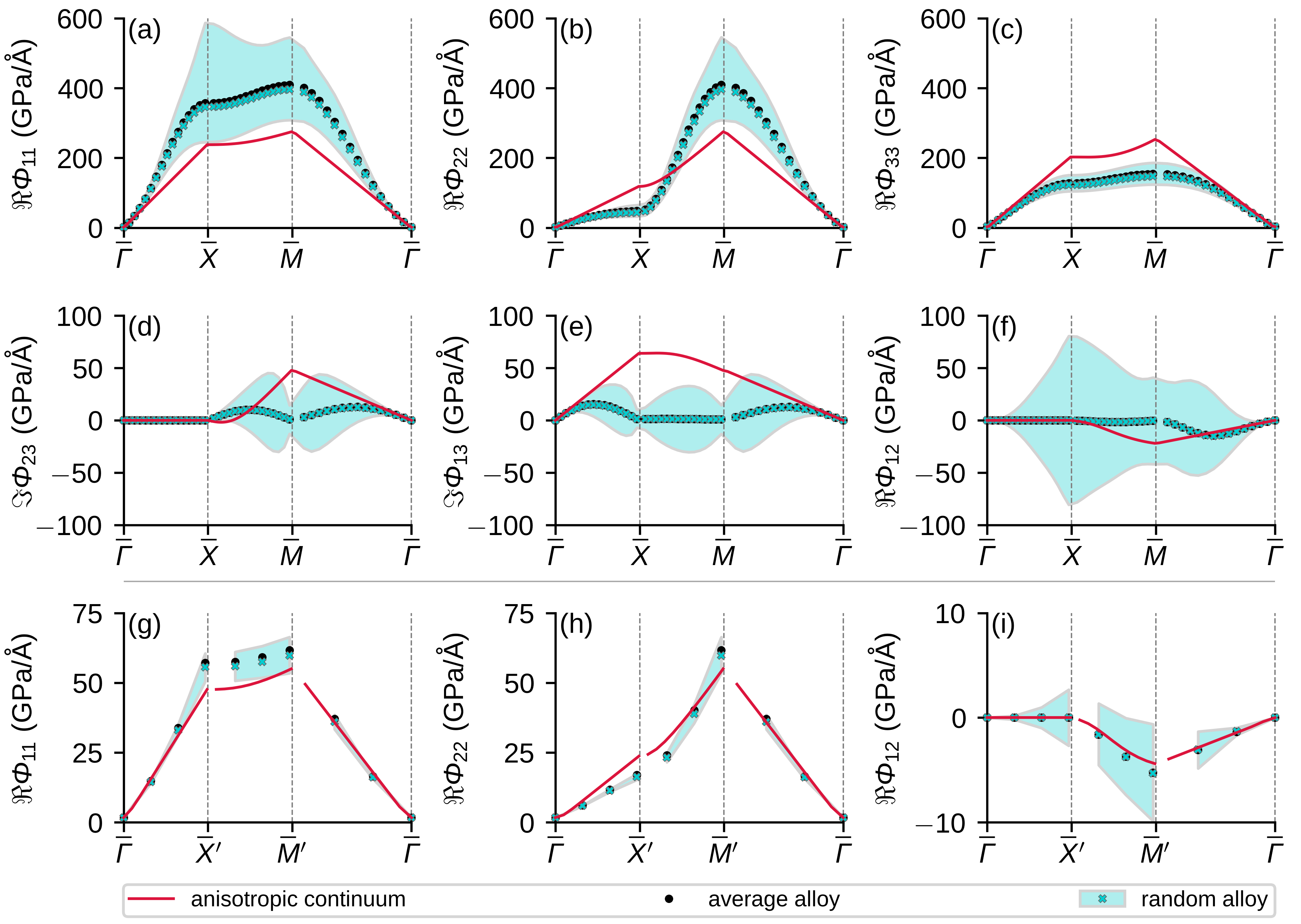}
    \caption{
    Surface stiffness of an equicomposition FeNiCr alloy;  the upper 
    two rows show the
    real parts of $\surfstiffcomp{11}$, $\surfstiffcomp{22}$,
$\surfstiffcomp{33}$, $\surfstiffcomp{12}$, as well as
the imaginary parts of $\surfstiffcomp{13}$ and
$\surfstiffcomp{23}$ along the path $\bar{\Gamma}$-$\bar{X}$-$\bar{M}$-$\bar{\Gamma}$ through the first quadrant 
    of the surface Brillouin zone (see \fig{fig:brillouin}); the stiffness
    of the average alloy is close to the effective mean stiffness of the true 
    random alloy according to \equ{equ:harmonicavg}; turquoise areas indicate the range between the 
    10th and 90th  percentile of the 
    distribution of site stiffnesses; 
    stiffness fluctuations are small at long wavelengths; with the 
    exception of $\surfstiffcomp{33}$, fluctuations grow significantly at 
    short wavelengths; the lower row shows the values
    of $\surfstiffcomp{11}$, $\surfstiffcomp{22}$, and $\surfstiffcomp{12}$ along the path $\bar{\Gamma}$-$\bar{X}'$-$\bar{M}'$-$\bar{\Gamma}$ through the long wavelength
    region; fluctuations in the random alloy, and the relative discrepancy between the atomistic data
    and the anisotropic-elastic model (red line) are smaller than
    at short wavelengths
    }
    \label{fig:result_alloy}
\end{figure*}

\fig{fig:result_alloy} shows
the stiffness components along the paths $\bar{\Gamma}$-$\bar{X}$-$\bar{M}$-$\bar{\Gamma}$ and $\bar{\Gamma}$-$\bar{X}'$-$\bar{M}'$-$\bar{\Gamma}$.
The upper two rows show the 
real parts of $\surfstiffcomp{11}$, $\surfstiffcomp{22}$,
$\surfstiffcomp{33}$, $\surfstiffcomp{12}$, as well as
the imaginary parts of $\surfstiffcomp{13}$ and
$\surfstiffcomp{23}$ along $\bar{\Gamma}$-$\bar{X}$-$\bar{M}$-$\bar{\Gamma}$. 
The bottom row shows $\Re\surfstiffcomp{11}$, $\Re\surfstiffcomp{22}$, and $\Re\surfstiffcomp{12}$ along the long-wavelength path $\bar{\Gamma}$-$\bar{X}'$-$\bar{M}'$-$\bar{\Gamma}$. The average alloy 
behaves like a pure metal, hence $\surfgreent\left(\functionqq\right)$
is the same for all surface sites, and there is one unique $3\times{}3$ stiffness
matrix $\surfstifft\left(\functionqq\right)$ for every point 
$\left(\functionqq\right)$ in the Brillouin zone. In the random alloy, 
by contrast,  $\surfgreent\left(\functionqq\right)$ fluctuates, 
therefore a different stiffness $\surfstifft\left(\functionqq\right)$  would be obtained for a different element distribution.  In \fig{fig:result_alloy}, markers indicate the harmonic mean $\surfstifft\left(\functionqq\right)$, and shaded areas the 10th and 90th percentiles of the site stiffnesses $\surfstifft^{(\nu)}\left(\functionqq\right)$.  Sample-by-sample 
variations of  $\surfstifft\left(\functionqq\right)$ are negligible.
The difference between the corresponding 10th and 90th percentiles is less than the size of the markers of  $\surfstifft\left(\functionqq\right)$ in the plot.

$\surfstifft\left(\functionqq\right)$ should be compared to the corresponding average alloy data and the continuum solution. All three models yield similar results in the 
long-wavelength limit. For example, the mean value of $\Re\surfstiffcomp{11}$ and $\Re\surfstiffcomp{22}$ at $\bar{\Gamma}$ is \SI{1.70}{\giga\pascal\per\angstrom}. The average-alloy value 
is \SI{1.74}{\giga\pascal\per\angstrom} and the continuum solution 
is \SI{1.70}{\giga\pascal\per\angstrom}. The continuum solution fails 
at short wavelengths, as was observed already in the Ni example. 
However, the average alloy data remain comparatively close to the mean values of the random alloy. For example, 
the relative difference between $\Re\surfstiffcomp{11}$ of the average and 
random alloy varies between \SI{3}{\percent} and \SI{4}{\percent} along the path. The continuum solution, on the other hand, underestimates the mean along most of the 
path, except near $\bar{\Gamma}$. Near $\bar{X}$, the continuum value is 
\SI{33}{\percent} lower than the mean value of the random alloy. 
The absolute value of the relative difference between $\Re\surfstiffcomp{22}$ of the average and random  alloy varies between \SI{3}{\percent} and \SI{11}{\percent}. In the case of $\Re\surfstiffcomp{33}$, the absolute relative difference does not exceed \SI{4}{\percent}.
Recall that the continuum solution for $\Re\surfstiffcomp{12}$ of pure Ni was qualitatively different from the atomistic solution. The same is true for the alloy. The continuum solution has a local minimum at $\bar{M}$, whereas both atomistic solutions have their
minimum between $\bar{M}$ and $\bar{\Gamma}$.

Fluctuations in the random alloy data are close to zero at $\bar{\Gamma}$, 
but increase with decreasing wavelength. For example, the relative difference between the 90th percentile of the site-specific values of $\Re\surfstiffcomp{11}$ and the mean according to \equ{equ:harmonicavg} increases from less than \SI{1}{\percent} of the 
mean value at $\bar{\Gamma}$ to \SI{70}{\percent}  at $\bar{X}$. 
Interestingly, it decreases to \SI{38}{\percent} at $\bar{M}$, 
even though $\bar{M}$ represents the limit of short wavelengths in both in-plane
directions. The maximum fluctuations of $\Re\surfstiffcomp{33}$ are smaller than those of $\Re\surfstiffcomp{11}$ and $\Re\surfstiffcomp{22}$. 

\begin{figure*}[t]
    \centering
    \includegraphics{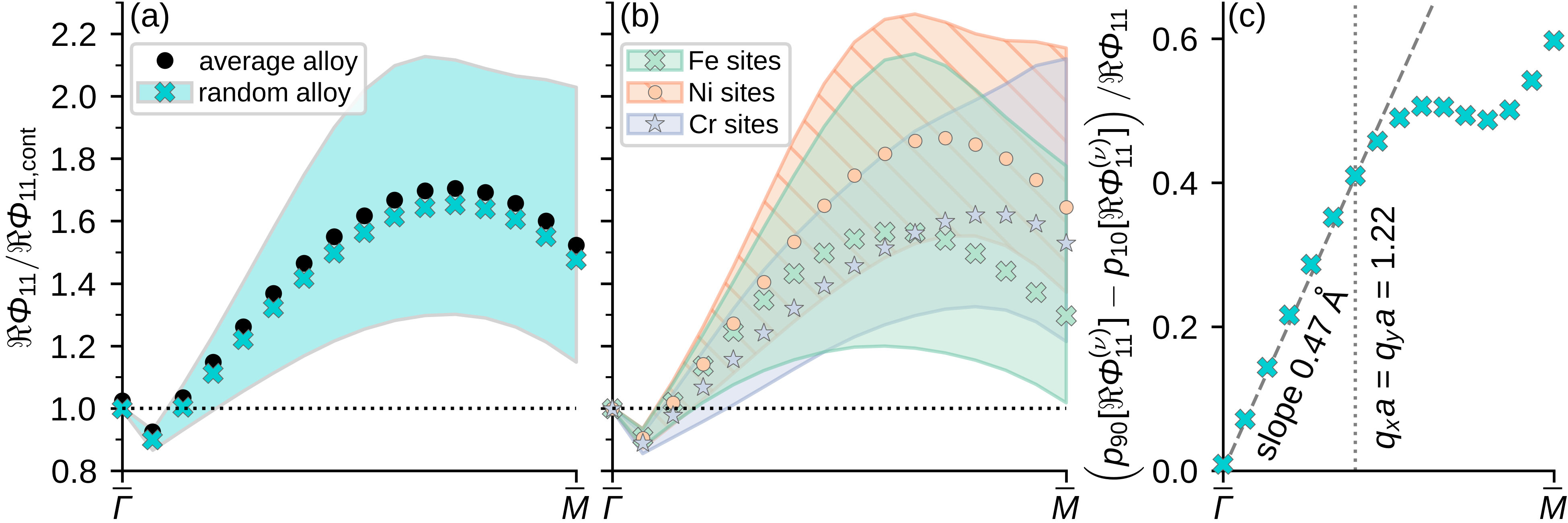}
    \caption{
    (a) Real part of $\surfstiffcomp{11}$ of average and true random FeNiCr, 
    divided by the continuum solution; values along the 
    diagonal $\bar{\Gamma}$-$\bar{M}$ of the first quadrant of the
    Brillouin zone; the shaded area shows the range between the 
    10th and 90th percentile of the 
    site stiffness distribution in the random alloy; 
    fluctuations in the random alloy go to zero as $\v{q}\rightarrow\bar{\Gamma}$;
    the mean value and the average alloy stiffness approach the 
    continuum stiffness; (b) (same y-axis as (a)) shows the mean value and the standard deviation of $\Phi^{(\nu)}_{11}$ in the random alloy for sites $\nu$ with different elements;  the fluctuations go to zero as $\v{q}\rightarrow\bar{\Gamma}$ and the mean values converge; (c) difference between the 90th  and 10th percentile of $\Phi^{(\nu)}_{11}$, divided by the harmonic mean; the dashed line 
    with a slope of \SI{0.47}{\angstrom} is the result of a linear least-squares fit between 
    $\bar{\Gamma}$ and $q_xa{}=q_ya=1.22$ (wavelength $\lambda_x = \lambda_y \approx{}5a$).
    }
    \label{fig:convergence_diagonal}
\end{figure*}

\fig{fig:convergence_diagonal} gives a more detailed view of 
convergence in the long wavelength limit, using the example of 
$\Re\surfstiffcomp{11}$. \fig{fig:convergence_diagonal}(a) shows the ratio 
between the real values of the atomistic and continuum data 
along the diagonal $\bar{\Gamma}$-$\bar{M}$. At $\bar{\Gamma}$,
the relative error between random alloy and continuum is  below  \SI{1}{\percent}. 
The average alloy stiffness is \SI{2}{\percent}
higher than the continuum value. The second point corresponds to the largest finite 
wavelength $(31a,31a)$. Here, both atomistic systems are softer
than the continuum. At wavelengths shorter than $(15.5a,15.5a)$, the 
atomistic systems are stiffer than the continuum. \fig{fig:convergence_diagonal}(b) distinguishes between contributions
from different elements. Let $\Omega_\mathrm{Fe}$ be the set of 
sites occupied by Fe atoms. To generate the corresponding curve
in \fig{fig:convergence_diagonal}(b), the average in 
\equ{equ:arithmeticavggreen} was restricted to sites $\nu\in\Omega_\mathrm{Fe}$.
Similarly, the 10th and 90th percentiles of $\surfstifft^{(\nu)}(\functionqq)$ were computed only for  this subset. The calculation for the other elements is analogous.  All three  mean values converge in the limit of long wavelengths to
the mean stiffness across all sites, and 
all fluctuations become minimal. 
Finally, in \fig{fig:convergence_diagonal}(c) we examined the growth
of the dispersion of  $\Re\surfstiffcomp{11}^{(\nu)}$ along  $\bar{\Gamma}$-$\bar{M}$. 
The figure shows the difference between 
the 90th and 10th percentile of  $\Re\surfstiffcomp{11}^{(\nu)}$, 
divided by the harmonic mean. 
Between $\bar{\Gamma}$ and $q_xa=q_ya=1.22$ (corresponding
to a wavelength of $\lambda_x=\lambda_y\approx{}5a$), the ratio grows approximately linearly. 
A linear least-squares fit yields a slope of \SI{0.47}{\angstrom}.

\section{Discussion}

We observed that the average alloy accurately approximates the 
mean surface stiffness of the true random alloy over the surface Brillouin zone. The normal  stiffness components $\surfstiffcomp{11}$, $\surfstiffcomp{22}$, and $\surfstiffcomp{33}$  are particularly important, because they 
generate the largest contributions to the force. The relative difference 
between the corresponding A-atom and and random alloy values is typically 
small. We observed the largest relative errors in $\surfstiffcomp{22}$ near $\bar{X}$, which represents the limit of short $\xc{1}$-wavelength and infinite $\xc{2}$-wavelength (and vice-versa by symmetry). However, the absolute value of $\surfstiffcomp{22}$ is also comparatively small between $\bar{\Gamma}$ and $\bar{X}$, yielding larger relative errors. $\surfstiffcomp{22}$ quickly increases
with decreasing wavelength in $\xc{2}$-direction and the relative
error decreases. $\surfstiffcomp{11}$ behaves similarly near the 
upper corner of the Brillouin zone. Given that the overall errors are small over the full surface Brillouin zone, we conclude that the average alloy provides an accurate estimate of the mean forces on atoms in the random. 
This result is not surprising,
since it was already shown in Ref.~\cite{Varvenne_2016}
that the A-atom has similar elastic constants as the 
corresponding true random alloy and the dependence of $\surfstifft$ near the Brillouin-zone edge is determined primarily by lattice structure.

Another question is when continuum elasticity becomes a good approximation. This question can only be answered with respect to a relative measure that quantifies what we mean by ``good''. Figure~\ref{fig:convergence_diagonal}(a) shows such a relative measure for one of the components of the stiffness tensor. The point at $\bar{\Gamma}$ and next to it are affected by the finite depth of the sample, as they represent homogeneous deformation and deformation with a wavelength equal to the size of our box. The next two points have an error of around $10\%$ and correspond to an $x$- and $y$-wavelength of $15.5 a$. It is fair to conclude that for distances beyond $\lambda_c\approx 15.5a$, the continuum approximation is reasonable. The decomposition into the individual atomic sites (Fig.~\ref{fig:convergence_diagonal}(b)) underlines this behavior, as the individual stiffnesses for Fe, Ni and Cr sites converge to a unified value near the point where the continuum solution appears appropriate.

A connected question is how the per-site fluctuations decay as a function of wavevector. It is clear that at large wavelength, where the continuum approximation holds, the per-site variation of $\surfstifft^{(\nu)}$ must be small. The per-site fluctuations are particularly small for the out-of-plane stiffness $\surfstiffcomp{33}$ (see Fig.~\ref{fig:result_alloy}(c)) where the continuum result has a lower error at intermediate wavelength as compared with the in-plane components $\surfstiffcomp{11}$ and $\surfstiffcomp{22}$ (Fig.~\ref{fig:result_alloy}(a)-(b)). The characteristic distance of $\lambda_c$ is therefore also a representative length beyond which per-site fluctuations become negligible. We now attempt a more quantitative analysis of this behavior. We divide the amplitude of the per-site fluctuations, as measured by the difference between the 90th and 10th percentile, by the stiffness matrix itself. As shown in Fig.~\ref{fig:convergence_diagonal}(c), this ratio depends linearly on the wavevector $q$. Since $\Phi\propto q$, this implies that the amplitude of per-site fluctuations decays with wavelength $\lambda$ as $\lambda^{-2}$. This decay of the fluctuations should be compared to the decay of the error in quantitative stochastic homogenization \cite{Gloria.2011, Gloria.2013vlu,MR3932093}, as shorter wavelengths are akin to taking smaller representative volumes of lateral length $\lambda$. Stochastic homogeneization predicts a scaling of $\lambda^{-2}$ for two-dimensional lattices.

Note that $\surfstiffcomp{33}$ is important for contact calculations~\cite{campana_practical_2006,pastewka_seamless_2012} while $\surfstiffcomp{11}$ and $\surfstiffcomp{22}$ are required for extended Peierls-Nabarro~\cite{Peierls_size,Nabarro_Simple_Cubic} models of dislocations~\cite{sharp_elasticity_2016,sharp_scale-_2017} or sliding friction~\cite{monti_sliding_2020}. For dislocations in high-entropy alloys it is therefore of particular importance to capture the by-site fluctuations appropriately. Our results show that the A-atom potential could be used for calculating the
\emph{mean} response. The per-site fluctuations could then be obtained from the local structure within regions of size $\sim\lambda_c$ around the site of interest using perturbative approaches~\cite{tewary_green-function_1973,tewary_lattice_1989,thomson_lattice_1992,tewary_lattice_1992,ohsawa_lattice_1996} or multipole expansions~\cite{bella_effective_2020}. Since a local evaluation up to a cutoff distance scales linearly with the number of sites, this would give rise to a feasible computational scheme.

As a model materials, we have considered only the case of a random ternary alloy in this study. Our method is not limited to the ternary case, and we do not expect qualitative changes in other random alloys. We speculate that a decrease of fluctuations can be expected in alloys with short range order, where site occupations are correlated. The A-atom method assumes entirely uncorrelated site occupations, so it may fail to accurately predict the mean alloy surface stiffness in this case. There is also a simple practical limitation when it comes to calculating the surface stiffness of other alloys: a suitable potential needs to be available. This potential should have a continuous and smooth cutoff such that the Hessian is well-defined (i.e., does not exhibit jumps). It is known that cutoff schemes can lead to spurious effects, for example when calculating the phonon spectrum \cite{mizuno_cutoff_2016,shimada_anomalous_2018}, which also implicitly relies on the Hessian.

\section{Summary and Outlook}

We have calculated the effective surface stiffness of unary crystals and a three-component random (high-entropy) alloys. Our results show that the surface stiffness has significant per-site variation near the edge of the surface Brillouin-zone, but that these variations disappear for larger wavelengths upon approach to the continuum limit. We identify a length of roughly $15.5$ atomic distances as the threshold where the continuum limit applies and per-site variations are small. At smaller distances, the average-atom approach of Varvenne and co-workers~\cite{Varvenne_2016} accurately captures the mean response of the solid. Our results are a first step towards building multi-scale Peierls-Nabarro type models for dislocations in high-entropy alloys, that require an accurate model for the elastic response of the crystalline material that encloses the dislocation. The next step is to derive a perturbative expansion around the mean-field results presented here that allows the efficient calculation of per-site surface stiffnesses.

\begin{acknowledgments}
We are grateful for many useful discussions with Mark Robbins, Tristan Sharp, Joseph Monti and Antoine Sanner. Simulations were carried out with \textsc{lammps}~\cite{plimpton_fast_1995} and \textsc{ase}~\cite{hjorth_larsen_atomic_2017}. Atomic configurations were rendered with \textsc{ovito}~\cite{stukowski_visualization_2009}. The authors acknowledge support from the Deutsche Forschungsgemeinschaft (grants PA 2023/4, DO 1412/4) and the European Research Council (grant 757343). Simulations were carried out at the J\"ulich Supercomputing Centre (grant hka18) and on NEMO at the University of Freiburg (DFG grant INST 39/963-1 FUGG).
\end{acknowledgments}

\appendix

\section{Lattice parameters and elastic constants\label{app:propertycalc}}

The \SI{0}{\kelvin} FCC lattice parameter and cubic elastic constants
of Ni, A-atom FeNiCr, and random FeNiCr are listed in table \tab{tab:material_properties}. 

\begin{table}[htb]
    \centering    
    \caption{\SI{0}{\kelvin} FCC lattice parameter and cubic elastic constants
    of pure Ni, FeNiCr equicomposition random alloy, and the corresponding
    average alloy, modeled using the EAM potential by Bonny et al.\ \cite{bonny_interatomic_2011}
    \label{tab:material_properties}}
    \begin{tabular}{cccc}
    \toprule
       & pure Ni & random alloy & average alloy \\ 
       \midrule
    $a_0$    (\SI{}{\angstrom})    
        & \SI[round-mode=places,round-precision=5]{3.519294473157436}{}   
        & \SI[round-mode=places,round-precision=5]{3.52137032484813}{} 
        & \SI[round-mode=places,round-precision=5]{3.521818615513773}{} \\
    $C_{11}$ (\SI{}{\giga\pascal}) 
        & \SI[round-mode=places,round-precision=2]{246.929958899558983}{} 
        & \SI[round-mode=places,round-precision=2]{243.39483263182854}{} 
        & \SI[round-mode=places,round-precision=2]{246.610145262896367}{}\\
    $C_{12}$ (\SI{}{\giga\pascal}) 
        & \SI[round-mode=places,round-precision=2]{147.071946812490665}{} 
        & \SI[round-mode=places,round-precision=2]{157.44603732988062}{} 
        & \SI[round-mode=places,round-precision=2]{158.121884686608013}{}\\
    $C_{44}$ (\SI{}{\giga\pascal}) 
        & \SI[round-mode=places,round-precision=2]{125.030567589943999}{} 
        & \SI[round-mode=places,round-precision=2]{134.9885847097457}{} 
        & \SI[round-mode=places,round-precision=2]{138.525095411635675}{}\\ 
    \bottomrule
    \end{tabular}
\end{table}

The properties of Ni and A-atom FeNiCr were calculated in the same way. In order to determine the \SI{0}{\kelvin} FCC lattice parameter, we minimized the pressure of a fully periodic  $5\times5\times5$ unit cell FCC crystal as a function of lattice parameter.  The residual pressure was less than \SI{1e-2}{\pascal} in both cases. The elastic constants were computed by imposing (\SI{1e-6}{}) strains  $\varepsilon_{ij}$ on a $5\times5\times5$ unit cell FCC crystal and measuring the stress response $\sigma_{ij}$.

In the case of random FeNiCr, we prepared three 
$30\times30\times30$ unit cell configurations  
with random site occupations. As before, we 
determined the \SI{0}{\kelvin} lattice parameter
by minimizing the pressure. However, the cell lengths
along $\xc{1}$, $\xc{2}$, and $\xc{3}$ were adjusted
independently, and so each calculation yields
three values for the lattice parameter. The value listed in \tab{tab:material_properties} is the average
over spatial directions and samples.
In order to determine the elastic constants, we 
imposed  simple shear strain $\varepsilon_{12}$ and uni-axial normal strain $\varepsilon_{22}$ on the samples and measured the stress response. In the
case of uniaxial tension/compression, Hooke's
law yields
\begin{align}
    \sigma_{22} &= C_{22} \varepsilon_{22}, \\
    \sigma_{11} &= C_{12} \varepsilon_{22}, \\
    \sigma_{33} &= C_{23} \varepsilon_{22},
\end{align}
where $C_{12}=C_{23}$ for cubic materials. 
In the case of simple shear 
\begin{align}
    \sigma_{12} = C_{44} \varepsilon_{12}.
\end{align}
We applied positive and negative shear, 
as well as tension and compression, with 
absolute values in the range \numrange{1e-9}{1e-4}.

\section{Hessian matrix within the embedded atom method \label{app:hessian}}

Below, we present the analytical solution for the Hessian matrix of 
the total potential energy in the EAM approximation. Greek superscripts refer to atom identifiers. For \natoms consecutively numbered atoms $\mu, \nu, \gamma, \delta \in [1,\natoms]$.  Lowercase roman subscripts refer to the components of a vector or tensor with respect to the three axes of a Cartesian coordinate system, i.e.\ $i,j\in[1,2,3]$. 
$\xc{i}[\nu]$ is the coordinate of atom $\nu$ in direction $i$. The $i$-component of the 
distance vector between atoms $\nu$ and $\mu$ is 
\begin{align}
    \xc{i}[\nu\mu{}] =\xc{i}[\mu{}]-\xc{i}[\nu{}].
\end{align}
The absolute value of the distance vector is 
\begin{align}
    \xabs{\nu\mu{}} &= \left(
     \left(\xc{1}[\nu\mu{}]\right)^2
    +\left(\xc{2}[\nu\mu{}]\right)^2
    +\left(\xc{3}[\nu\mu{}]\right)^2
    \right)^{1/2}.
\end{align}
Furthermore, we use the abbreviations
\begin{align}
\xcn{i}[\nu\mu{}]    &\equiv \frac{\xc{i}[\nu\mu{}]}{\xabs{\nu\mu}} \quad\text{and} \\
\xco{ij}[\nu{}\mu{}] &\equiv \xcn{i}[\nu\mu{}]\xcn{j}[\nu\mu{}].
\end{align}
The two symbols represent a normalized distance vector and the outer product
of a normalized distance vector with itself, respectively. The expression for the Hessian involves the following derivatives of a pair distance vector and its absolute value: 
\begin{align}
    \frac{\partial \xc{i}[\gamma\delta{}]}{\partial \xc{j}[\mu{}]} &= \delta_{ij}\left(\delta_{\mu\delta}-\delta_{\mu\gamma}\right), \\
    \frac{\partial \xabs{\gamma\delta}}{\partial \xc{i}[\nu]} &= \xcn{i}[\gamma{}\delta{}] \left(\delta_{\delta\nu}-\delta_{\gamma\nu}\right),
\end{align}
where $\delta_{ij}$ is Kronecker's delta, i.e.\ $\delta_{ij}=1$ if $i=j$ and $\delta_{ij}=0$ otherwise.

In the EAM, the total potential energy $\mathcal{V}^\mathrm{int}$ due to interaction between atoms is the sum of pair and embedding energy contributions \cite{Daw_1984},
\begin{align}
            \mathcal{V}^\mathrm{int} 
        &= \mathcal{V}^\mathrm{pair} + \mathcal{V}^\mathrm{embed}.
\end{align}
The pair energy contribution is 
\begin{align}
        \mathcal{V}^\mathrm{pair} 
        &= \frac{1}{2}\sum_{\gamma}^{\natoms}\sum_{\delta\neq\gamma}^{\natoms}\pairpot{\gamma\delta}\left(\xabs{\gamma\delta}\right),
\end{align}
where $\pairpot{\nu\mu}\left(\xabs{\gamma\delta}\right)$ is the pair potential of atoms  $\mu$ and $\nu$, evaluated at $\xabs{\gamma\delta}$. For the sake of brevity, we use the  abbreviation $\phi^{(\gamma\delta)} \equiv \pairpot{\gamma\delta}\left(\xabs{\gamma\delta}\right)$
in the following.

The embedding energy contribution is 
\begin{align}
        \mathcal{V}^\mathrm{embed}
        = \sum_{\gamma}^{\natoms}\embedfun{\gamma}\left(\totedens{\gamma}\right),
\end{align}
where $\embedfun{\gamma}$ is the embedding energy functional of atom $\gamma$. 
 $\embedfun{\gamma}$ is a functional of the total electron density at the site of $\gamma$, which is computed as 
 \begin{align}        
        \totedens{\gamma} &= \sum_{\delta\neq\gamma}^{\natoms}  \edensfun{\delta}\left(\xabs{\gamma\delta}\right),
\end{align}
where $\edensfun{\delta}\left(\xabs{\gamma\delta}\right)$ is the electron density function of atom $\delta$. For the sake of brevity, we write
 $\edensfun{\gamma\delta}\equiv \edensfun{\delta}\left(\xabs{\gamma\delta}\right)$.

The Hessian matrix $\mathbf{H}$ is the $3\natoms\times{}3\natoms$ matrix of second derivatives of $\mathcal{V}^\mathrm{int}$ with respect to the coordinates of the atoms. The components of $\mathbf{H}$ are
\begin{align}
H_{(3(\nu{}-1)+i)(3(\mu{}-1)+j)} = \frac{\partial^2 \mathcal{V}^\mathrm{int}}{\partial \xc{i}[\nu{}]\partial \xc{j}[\mu{}]}.
\end{align}

We first write the gradient of $\mathcal{V}^\mathrm{int}$. In the following, we use one and two dashes, respectively, to indicate the first and second derivatives of a function, e.g.\ ${\edensfun{\delta}}'\left(\xabs{\gamma\delta}\right) \equiv d{\edensfun{\delta}}(\xabs{\gamma\delta})/d\xabs{\gamma\delta}$ and ${\edensfun{\delta}}''\left(\xabs{\gamma\delta}\right) \equiv d^2\edensfun{\delta}(\xabs{\gamma\delta})/d(\xabs{\gamma\delta})^2$. As before, we abbreviate the dependence on $\xabs{\gamma\delta}$ by writing  ${\edensfun{\gamma\delta}}
'\equiv {\edensfun{\delta}}'\left(\xabs{\gamma\delta}\right)$ and ${\pairpot{\gamma\delta}}' \equiv {\pairpot{\gamma\delta}}'\left(\xabs{\gamma\delta}\right)$, and likewise for the second 
derivative. With this notation, the expression for the gradient of  $\mathcal{V}^\mathrm{int}$ becomes
\begin{align}
\begin{aligned}
\frac{\partial \mathcal{V}^\mathrm{int}}{\partial   \xc{i}[\nu]} &= 
\frac{\partial \mathcal{V}^\mathrm{pair}}{ \partial \xc{i}[\nu]} +
\frac{\partial \mathcal{V}^\mathrm{embed}}{\partial \xc{i}[\nu]}  \\
&=
-\sum_{\gamma\neq\nu}^{\natoms}
\left(\pairpot{\gamma\nu}' + \embedfun{\gamma}' \edensfun{\gamma\nu}' + \embedfun{\nu}'\edensfun{\nu\gamma}'
\right)\xcn{i}[\nu\gamma].
\end{aligned}
\end{align}

Like the gradient, the Hessian matrix can be split into contributions from 
$\mathcal{V}^\mathrm{pair}$ and $\mathcal{V}^\mathrm{embed}$, 
\begin{align}
\frac{\partial^2   \mathcal{V}^\mathrm{int}}{\partial \xc{i}[\nu]\partial \xc{j}[\mu]} = 
\frac{\partial^2 \mathcal{V}^\mathrm{pair}}{ \partial \xc{i}[\nu]\partial \xc{j}[\mu]} +
\frac{\partial^2 \mathcal{V}^\mathrm{embed}}{\partial \xc{i}[\nu]\partial \xc{j}[\mu]}.    
\end{align}
The pair contribution is 
\begin{align}
\begin{aligned}
\frac{\partial^2 \mathcal{V}^\mathrm{pair}}{ \partial \xc{i}[\nu] \partial \xc{j}[\mu]} = 
&-\pairpot{\nu\mu}'' 
\xco{ij}[\nu{}\mu{}]
-\frac{\pairpot{\nu\mu}'}{\xabs{\nu\mu}}\left(
\delta_{ij}- \xco{ij}[\nu{}\mu{}]
\right) \\ 
&+\delta_{\nu\mu}\sum_{\gamma\neq\nu}^{\natoms}
\pairpot{\nu\gamma}'' \xco{ij}[\nu{}\gamma{}]\\
&+\delta_{\nu\mu}\sum_{\gamma\neq\nu}^{\natoms}\frac{\pairpot{\nu\gamma}'}{\xabs{\nu\gamma}}\left(
\delta_{ij}-
\xco{ij}[\nu{}\gamma{}]
\right).
\end{aligned}
\end{align}
The third and the fourth term are the sums of the first and second term, respectively, over the neighbors of $\nu$. This expression is equal to the 
Hessian matrix for a pair potential, see Ref.~\cite{pastewka_seamless_2012}.

The embedding contribution is the sum of eight terms,
\begin{align}
\frac{\partial^2 \mathcal{V}^\mathrm{embed}}{\partial \xc{i}[\nu] \partial \xc{j}[\mu]} 
        &= \sum_{n=1}^{8}\hessianterm{n},
\end{align}
where
\begin{align}
\hessianterm{1}&= 
\delta_{\nu\mu}\embedfun{\nu}''
\sum_{\gamma\neq\nu}^{\natoms}\edensfun{\nu\gamma}'\xc{i}[\nu\gamma{}]
\sum_{\gamma\neq\nu}^{\natoms}\edensfun{\nu\gamma}'\xc{j}[\nu\gamma{}],
\end{align}
\begin{align}
\hessianterm{2} &= 
-\embedfun{\nu}''\edensfun{\nu\mu}' \xc{j}[\nu\mu] \sum_{\gamma\neq\nu}^{\natoms} 
\edensfun{\nu\gamma}' \xc{i}[\nu\gamma], 
\end{align}
\begin{align}
\hessianterm{3}&=
\embedfun{\mu}''\edensfun{\mu\nu}' \xc{i}[\nu\mu{}] \sum_{\gamma\neq\mu}^{\natoms} 
\edensfun{\mu\gamma}' \xc{j}[\mu\gamma{}], 
\end{align}
\begin{align}
\hessianterm{4} &= -\left(\embedfun{\mu}'\edensfun{\mu\nu}'' + \embedfun{\nu}'\edensfun{\nu\mu}''\right)
\xco{ij}[\nu{}\mu{}],
\end{align}
\begin{align}
\hessianterm{5} &= \delta_{\nu\mu} \sum_{\gamma\neq\nu}^{\natoms}
\left(\embedfun{\gamma}'\edensfun{\gamma\nu}'' + \embedfun{\nu}'\edensfun{\nu\gamma}''\right)
\xco{i{}j}[\nu{}\gamma{}],
\end{align}
\begin{align}
\hessianterm{6}&= -\frac{\embedfun{\mu}'\edensfun{\mu\nu}' + \embedfun{\nu}'\edensfun{\nu\mu}'}{\xabs{\nu\mu}}
\left(
\delta_{ij}- 
\xco{i{}j}[\nu{}\mu{}]
\right),\
\end{align}
\begin{align}
\hessianterm{7}&= \delta_{\nu\mu} 
\sum_{\gamma\neq\nu}^{\natoms}
\frac{\embedfun{\gamma}'\edensfun{\gamma\nu}' + \embedfun{\nu}'\edensfun{\nu\gamma}'}{\xabs{\nu\gamma}}
\left(\delta_{ij}-\xco{ij}[\nu{}\gamma{}]\right),
\end{align}
and
\begin{align}
\hessianterm{8}&= \sum_{\substack{\gamma\neq\nu \\ \gamma \neq \mu}}^{\natoms}
\embedfun{\gamma}'' \edensfun{\gamma\nu}'\edensfun{\gamma\mu}' 
\frac{\xc{i}[\gamma\nu]}{\xabs{\gamma\nu}}
\frac{\xc{j}[\gamma\mu]}{\xabs{\gamma\mu}}.
\end{align}
Note that the terms remain the same when the index pairs $(i,\nu)$ and $(j,\mu)$ are interchanged, which is necessary for the Hessian to be symmetric. Terms $\hessianterm{5}$ and $\hessianterm{7}$ are the sums of terms $\hessianterm{4}$ and $\hessianterm{6}$, respectively, over the neighbors of atom $\nu$. Terms $\hessianterm{1}$--$\hessianterm{7}$ are zero if atoms $\mu$ and $\nu$ are not neighbors, i.e.\ if $r_{\mu\nu}>r_{\mathrm{cut}}$, where $r_{\mathrm{cut}}$ is 
the cutoff radius of the potential. Term $\hessianterm{8}$ is the most complex term. In order to compute this term, one needs to determine the common neighbors of atoms $\mu$ and $\nu$, even if $\mu$ and $\nu$ are not neighbors themselves.

\section{Surface Green's function for an anisotropic elastic continuum of finite thickness \label{app:conti}}

No closed form solution for the surface Green's function of the anistropic elastic continuum exists. We here compute this Green's function semi-analytically. Starting from the definition of the (small-strain) strain tensor,
\begin{align}       
    \varepsilon_{ij} = \frac{1}{2}\left(\partial_i u_j + \partial_j u_i\right)
    \label{eq:strain}
\end{align}
where $\v{u}(\functionxxx)$ is the displacement field, we obtain the stress tensor as
\begin{align}
    \sigma_{ij} = C_{ijkl} \varepsilon_{kl} = \frac{1}{2} C_{ijkl} \left(\partial_k u_l + \partial_l u_k\right) = C_{ijkl} \partial_k u_l.
    \label{eq:Hooke}
\end{align}
$C_{ijkl}$ is the fourth-order tensor of elastic constants with at most $27$ independent component. Note that $\partial_i$ indicates the partial derivative in direction $i$ and Einstein summation convention applies to all Latin indices in this paper. The usual symmetry relationships
\begin{align}
    C_{ijkl}=C_{jikl}\text{,}\quad C_{ijkl}=C_{ijlk}\quad\text{and}\quad C_{ijkl}=C_{klij}
    \label{eq:sym}
\end{align}
have been used to obtain the last equality in Eq.~\eqref{eq:Hooke}.
Elastostatic equilibrium dictates $\partial_i \sigma_{ij}=0$. Inserting Eqs.~\eqref{eq:Hooke} into this expression yield the
generalization of the Navier-Lam\'e equations,
\begin{align}
  \partial_j\left(C_{ijkl} \partial_k u_l\right) = C_{ijkl} \partial_j \partial_k u_l = 0,
  \label{eq:elastostatic}
\end{align}
where we have assumed that $C_{ijkl}$ does not depend on position, i.e.\ we are dealing with a homogeneous half-space.

We now search for a solution of the displacements $\vec{u}_s(\functionxx)$ within the plane of a surface subject to the traction boundary conditions $P(\functionxx)$, $Q_1(\functionxx)$ and $Q_2(\functionxx)$. With $\v{\traction}=(Q_1,Q_2,P)$, the displacements are given by $\v{u}_s(\functionxx)=\int d\xc{1}'d\xc{2}'\, \surfgreent(\xc{1}-\xc{1}', \xc{2}-\xc{2}')\cdot\v{\traction}(\xc{1}',\xc{2}')$. It is usually convenient to state the Fourier transform of this expression, $\v{u}_s(\v{q})=\surfgreent(\v{q})\cdot\v{\traction}(\v{q})$ where $\v{q}=(\functionqq)$ is the wavevector within the plane of the surface. $\surfgreent$ is the surface Green's function.

Note that Eq.~\eqref{eq:elastostatic} is a set of three linear partial differential equations for the three components of the displacement field $\vec{u}(\functionxxx)$ throughout the body. We are only interested in $\vec{u}_s(\functionxx)=\vec{u}(\functionxx,0)$. 
We can write Eq.~\eqref{eq:elastostatic} as~\cite{chen_galerkin_1996}
\begin{align}
    M_{il} u_l = 0
    \label{eq:Mop}
\end{align}
with the linear operator
\begin{align}
    M_{il} = C_{ijkl}\partial_j\partial_k.
    \label{eq:Mopdef}
\end{align}
Because of Eq.~\eqref{eq:sym}, the operator $\t{M}$ is symmetric, $M_{il}=M_{li}$.
In order to obtain the surface Green's function, we need to impose the traction boundary conditions $P$, $Q_1$ and $Q_2$. At the surface ($\xc{3}=0$), the stress tensor fulfills
\begin{align}
\begin{aligned}
  \sigma_{33}(\xc{1},\xc{2},\xc{3}=0)&=P(\functionxx)
  \text{,} \\ 
  \sigma_{13}(\xc{1},\xc{2},\xc{3}=0)&=Q_1(\functionxx)\quad
  \text{and} \\ 
  \sigma_{23}(\xc{1},\xc{2},\xc{3}=0)&=Q_2(\functionxx).
  \label{eq:boundarycond}
\end{aligned}
\end{align}

To solve Eq.~\eqref{eq:Mop} numerically under the boundary conditions given by Eq.~\eqref{eq:boundarycond}, we need to transform Eq.~\eqref{eq:Mopdef} into an algebraic equation.
Because we are interested in the solution for a plane interface, we have translational invariance in the $\xc{1}$-$\xc{2}$ plane.
The Fourier transform of Eq.~\eqref{eq:Mopdef} in this plane is given by
\begin{align}
\begin{aligned}
    M_{il} = &- C_{i11l} {\qc{1}}^2 - C_{i22l} {\qc{2}}^2 - (C_{i12l}+C_{i21l}) \qc{1} \qc{2} \\
    &+ \left[i(C_{i13l} + C_{i31l}) \qc{1} + i(C_{i23l}+C_{i32l}) \qc{2}\right] \partial_{{3}} \\
    & + C_{i33l} {\partial_{{3}}}^2,
\end{aligned}
\end{align}
where $\qc{1}$ and $\qc{2}$ are the wavevectors in this plane. Any nontrivial solution to the homogeneous equation Eq.~\eqref{eq:Mop} must fulfill $\det \t{M}=0$. This fixes the admissible values of the eigenvalue $i\qc{3}$ of the operator $\partial_{{3}}$.
Since $\det\t{M}$ is a sixth-order even polynomial in $i\qc{3}$, for each $\qc{1},\qc{2}$, we obtain six values for $i{\qc{3}}^{(\alpha)}$ that occur in symmetric pairs. 

For six eigenvalues ${\qc{3}}^{(\alpha)}$ the displacement field is given by a superposition of the basis functions $\v{\eta}^{(\alpha)} e^{i{\qc{3}}^{(\alpha)}\xc{3}}$, where $\v{\eta}^{(\alpha)}$ is the solution of $\t{M}\cdot\v{\eta}^{(\alpha)} e^{i{\qc{3}}^{(\alpha)}\xc{3}}=0$. It is straightforward solve for both ${\qc{3}}^{(\alpha)}$ and $\v{\eta}^{(\alpha)}$ numerically. The general displacement field is then given by
\begin{align}
    \v{u} = \sum_\alpha c_\alpha \v{\eta}^{(\alpha)} e^{i\qc{3}^{(\alpha)}\xc{3}} = \t{U}(\qc{1},\qc{2},\xc{3})\cdot\vec{c}
    \label{eq:gendispl}
\end{align}
with generally $\v{c}=(c_1, c_2, c_3, c_4, c_5, c_6)$ and $U_{k\alpha}(\qc{1},\qc{2},\xc{3})=\eta_k^{(\alpha)} e^{i{\qc{3}}^{(\alpha)}\xc{3}}$ where $\eta_k^{(\alpha)}$ and ${\qc{3}}^{(\alpha)}$ depend implicitly on $\qc{1}$ and $\qc{2}$.
The constants $c_\alpha$ are now obtained from the displacement or traction boundary conditions on both top and bottom of the half-space.

For an infinite half-space, all $c_\alpha$ for $\Im \qc{3}^{(\alpha)}<0$ must vanish because the solution diverges as $\qc{3}\to\infty$. This leaves us with three relevant basis functions, that we label (without loss of generality) by $\alpha=1,2,3$ (and hence $c_4=c_5=c_6=0$). The traction boundary condition, Eq.~\eqref{eq:boundarycond}, becomes  
\begin{align}
Q_1 (\qc{1},\qc{2}) &= iC_{13kl} q_k u_l = i\sum_\alpha C_{13kl} q_k^{(\alpha)} \eta_l^{(\alpha)} c_\alpha
  \label{eq:bc1}\\
  Q_2 (\qc{1},\qc{2}) &= iC_{23kl} q_k u_l = i\sum_\alpha C_{23kl} q_k^{(\alpha)} \eta_l^{(\alpha)} c_\alpha,
  \label{eq:bc2}\\
  P (\qc{1},\qc{2}) &= iC_{33kl} q_k u_l = i\sum_\alpha C_{33kl} q_k^{(\alpha)} \eta_l^{(\alpha)} c_\alpha
  \label{eq:bc3}
  \end{align}
or in matrix notation
\begin{align}
  \v{\traction}(\qc{1},\qc{2}) = \t{F}(\qc{1},\qc{2})\cdot \v{c}
  \label{eq:surftrac}
\end{align}
with $F_{j\alpha}(\qc{1},\qc{2})=iC_{j\textbf{}3kl} q_k^{(\alpha)} \eta_l^{(\alpha)}$
and ${\qc{1}}^{(\alpha)}\equiv \qc{1}$ and ${\qc{2}}^{(\alpha)}\equiv \qc{2}$. Combining Eqs.~\eqref{eq:gendispl} and \eqref{eq:surftrac} gives $\surfgreent(\qc{1},\qc{2})=\t{U}(\qc{1},\qc{2},0)\cdot \t{F}^{-1}(\qc{1},\qc{2})$ or $\surfstifft=\surfgreent^{-1}=\t{F}\cdot\t{U}^{-1}$.

For a finite half-space, we need to keep all six basis functions $\alpha\in[1\ldots 6]$ and require in addition a boundary condition at the bottom of the substrate. We here only discuss fixed displacement, in particular $u_i(\xc{1},\xc{2},\xc{3}=h)=0$ where $h$ is the thickness of the elastic substrate. In addition to Eqs.~\eqref{eq:bc1} to \eqref{eq:bc3}, the displacement boundary condition leads to the additional equations
\begin{align}
    \sum_\alpha c_\alpha \eta_j^{(\alpha)} e^{i{\qc{3}}^{(\alpha)}h} = u_{b,j}
  \label{eq:bc4}
\end{align}
for $j=1,2,3$. Here $\v{u}_b$ are the displacements at the bottom of the substrate. In dyadic notation this becomes
\begin{align}
  \v{\varphi}(\qc{1},\qc{2}) = \t{F}_h(\qc{1},\qc{2})\cdot \v{c}
  \label{eq:surftrac2}
\end{align}
where $\v{\varphi}$ contains forces at the top and displacements at the bottom of the substrate. $\t{F}_h$ is a $6\times 6$ matrix. The Green's function is then given by the first three columns of $\t{U}\cdot \t{F}_h^{-1}$.

Note that for the (special) isotropic case where $C_{ijkl}=\lambda \delta_{ij}\delta_{kl}+2\mu \delta_{ik}\delta_{jl}$ with Lam\'e constants $\lambda$ and $\mu$, the solution of $\det\t{M}=0$ is degenerate, $i\qc{3}=\pm \sqrt{\qc{1}^2 + \qc{2}^2}$, and the above analysis does not apply. The displacement field is then given by superposition of the basis functions $e^{i\qc{3}\xc{3}}$, $\xc{3}e^{i\qc{3}\xc{3}}$, ${\xc{3}}^2e^{i\qc{3}\xc{3}}$, $e^{-i\qc{3}\xc{3}}$, $\xc{3}e^{-i\qc{3}\xc{3}}$ and ${\xc{3}}^2e^{-i\qc{3}\xc{3}}$. The close-form solution for the infinite half-space in this limit is described in Ref.~\cite{amba-rao_fourier_1969}. It yields the surface Green's function
\begin{align}
  \mu \surfgreent(\v{q}) = \begin{pmatrix}
    \frac{1}{q} - \frac{\nu {\qc{1}}^2}{q^3} & -\frac{\nu \qc{1}\qc{2}}{q^3} & i\frac{(1-2\nu) \qc{1}}{2 q^2} \\
    -\frac{\nu \qc{1}\qc{2}}{q^3} & \frac{1}{q} - \frac{\nu {\qc{2}}^2}{q^3} & i\frac{(1-2\nu) \qc{2}}{2 q^2} \\
    -i\frac{(1-2\nu) \qc{1}}{2 q^2} & -i\frac{(1-2\nu) \qc{2}}{2 q^2} & \frac{1-\nu}{q} \\
  \end{pmatrix}
  \label{eq:gfiso}
\end{align}
with Poisson number $\nu=\lambda/[2(\lambda+\mu)]$. Inverse Fourier transform of Eq.~\eqref{eq:gfiso} leads to the well-known potential functions of Boussinesq \& Cerruti~\cite{johnson_contact_1985}.

\bibliography{gf}

\end{document}